\documentclass[twocolumn,showpacs,amsfonts,amssymb,floats]{revtex4}
\usepackage{graphicx}% Include figure files
\usepackage{dcolumn}% Align table columns on decimal point
\usepackage{bm}% bold math

\begin{document}

\title{Energy landscape and shear modulus of interlayer Josephson vortex systems}
\author{Yoshihiko Nonomura}
\email{nonomura.yoshihiko@nims.go.jp}
\author{Xiao Hu}%
\affiliation{
Computational Materials Science Center, 
National Institute for Materials Science, 
Tsukuba, Ibaraki 305-0047, Japan
}
\date{\today}

\begin{abstract}
The ground state of interlayer Josephson vortex systems is investigated 
on the basis of a simplified Lawrence-Doniach model in which spatial 
dependence of the gauge field and the amplitude of superconducting order 
parameter is not taken into account. Energy landscape is drawn with 
respect to the in-plane field, the period of insulating layers including 
Josephson vortices, and the shift from the aligned vortex lattice. 
The energy landscape has a multi-valley structure and ground-state 
configurations correspond to bifurcation points of the valleys. In the 
high-field region, the shear modulus becomes independent of field and 
its anisotropy dependence is given by $c_{66}\propto \gamma^{-4}$.
\end{abstract}

\maketitle

{\it Introduction.}
Although interlayer Josephson vortex systems in cuprate high-$T_{\rm c}$ 
superconductors have been intensively studied, their phase diagrams 
have not been established yet even in the ground state. For high 
in-plane fields, Josephson vortices penetrate into every insulating 
layer, and the ground state is given by an elongated triangular 
lattice aligned along the superconducting layers. As the field 
decreases, such an aligned vortex lattice becomes unstable 
owing to the shear instability \cite{Ivlev}, and the shearing 
angle of vortex lattices increases continuously \cite{Koshelev06}. 
As the field further decreases, the rotated vortex lattices have been 
considered to be the ground state \cite{Campbell,Levitov,Lagna}. 
The studies mentioned above (except for Ref.\ \cite{Koshelev06}) 
were based on the London model, and effects of the layered 
structure were included only as geometrical constraint.

This layered structure can be directly taken into account by using the 
Lawrence-Doniach model. Bulaevskii and Clem \cite{Bulaevskii91} first 
introduced this model in order to investigate interlayer Josephson 
vortex systems for high fields. Ichioka \cite{Ichioka95} calculated the 
energies of the aligned vortex lattices including vacant insulating 
layers. Stability of some rotated-vortex-lattice configurations was 
first pointed out by Hu and Tachiki \cite{Hu98} in the frustrated 
XY model, and Ikeda \cite{Ikeda02} reported similar results in the 
Lawrence-Doniach model within the lowest-Landau-level approximation. 
Quite recently, Koshelev \cite{Koshelev06} systematically evaluated 
the ground-state phase diagram (including the sheared and rotated 
vortex lattices) based on a simplified Lawrence-Doniach model, 
in which spatial dependence of the gauge field and the amplitude 
of superconducting order parameter is neglected. 

The ground state of this simplified model is completely given by the 
in-plane field $h$, the period $N$ between insulating layers including 
Josephson vortices, and the shift $\Delta$ of the vortex lattice from 
the aligned one. In Ref.\ \cite{Koshelev06}, $\Delta$ is fixed at 
fractional numbers which correspond to the rigid rotated vortex 
lattices. In the present study, we draw the full energy landscape 
for the first time, and find that the energy landscape has a 
multi-valley structure, and that the ground state for low fields 
shows systematic deviation from the rigid rotated vortex lattices.
We also find that field dependence of the shear modulus is 
quite different from previous estimation based on the London 
theory \cite{Ivlev} in the high-field region.
\medskip
\par
\begin{figure}
\begin{center}
\includegraphics[width=42mm]{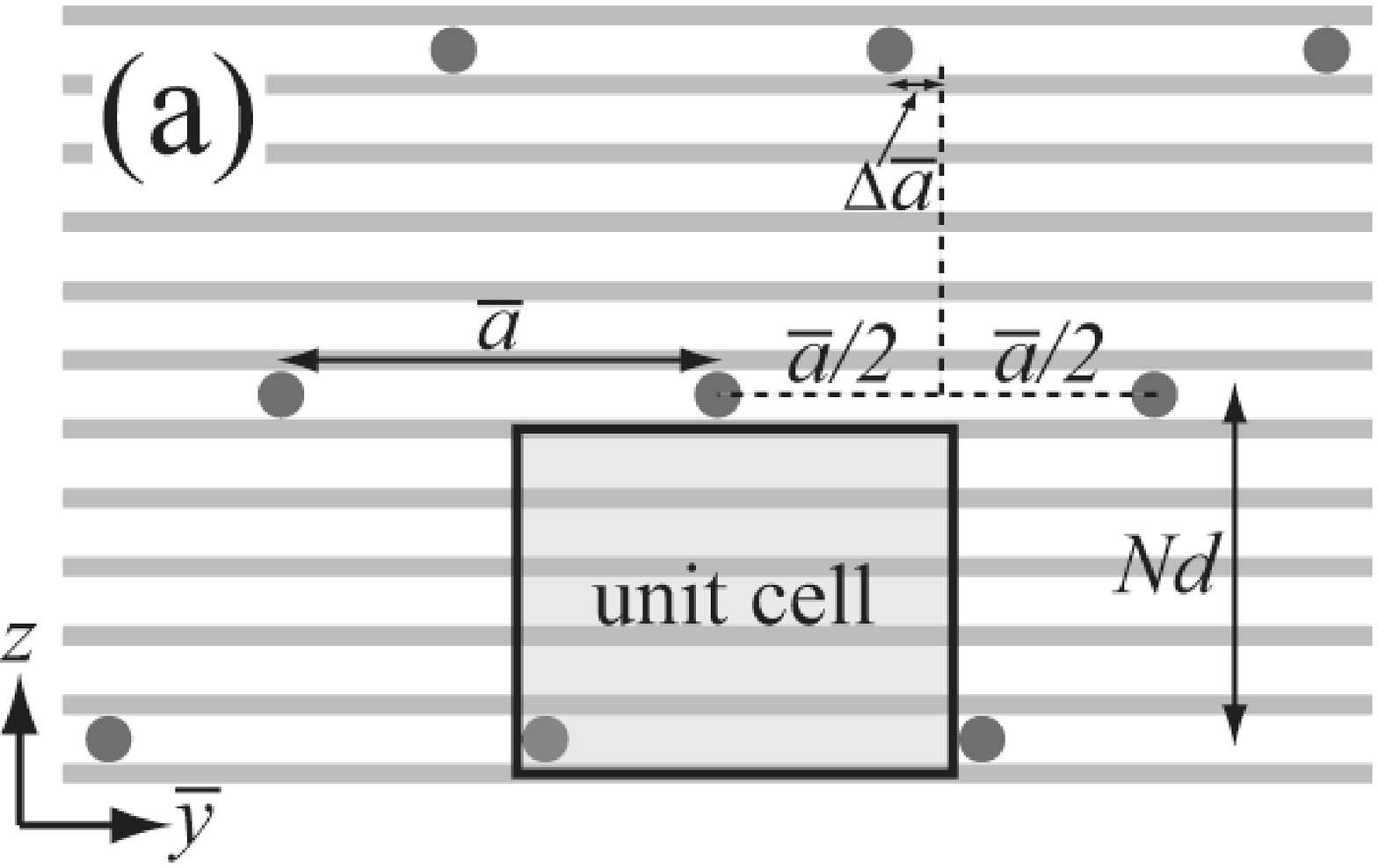}
\includegraphics[width=42mm]{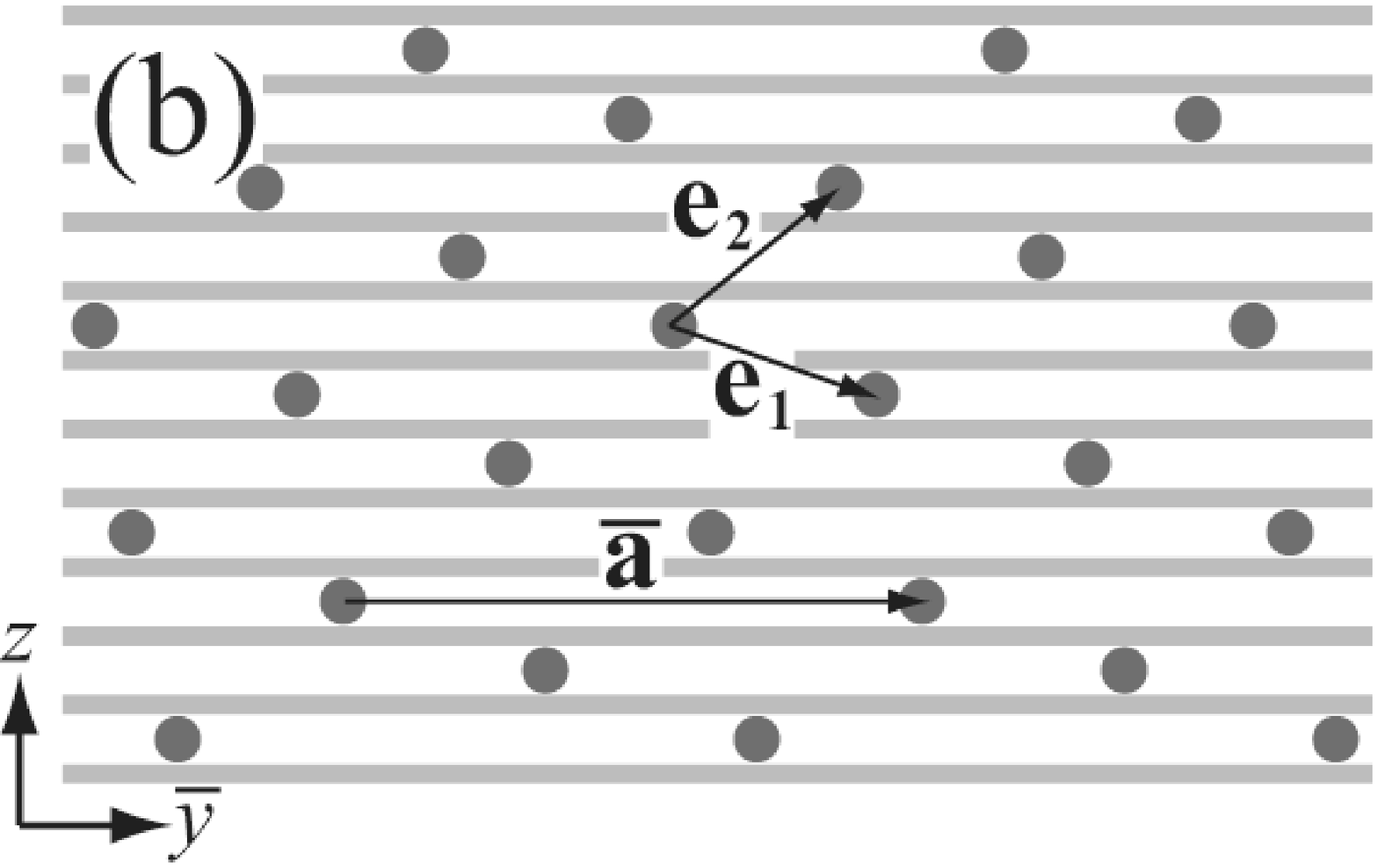}
\caption{\label{fig1}Definition of (a) the unit cell of a vortex 
lattice and (b) the labeling of a rigid rotated vortex lattice (here 
$(n,m,N)=(2,1,1)$). Scaled coordinate $\bar{y}\equiv y/(\gamma d)$ is used.}
\end{center}
\end{figure}

{\it Formulation.}
When spatial dependence of the gauge field and the 
amplitude of superconducting order parameter is neglected, 
the Lawrence-Doniach free energy functional is expressed 
by the phase component $\phi_{n}$ \cite{Koshelev06} as
\begin{eqnarray}
\label{simLD}
&&F\{\phi_{n}({\bf r})\}
=\frac{\phi_{0}^{2}}{16\pi^{3}\lambda_{ab}^{2}}
\int d^{3}{\bf r}\left\{\frac{1}{2}
\left({\bf \nabla} \phi_{n}\right)^{2}\right.\nonumber\\*
&&+\left.\frac{1}{(\gamma d)^{2}}\left[1-\cos
\left(\phi_{n+1}-\phi_{n}-\frac{2\pi}{\phi_{0}}d B_{x}y
      \right)\right]\right\},
\end{eqnarray}
where $\phi_{0}$ is the flux quantum, $\lambda_{ab}$ is 
the penetration depth in the $ab$ plane, $\gamma$ is the 
anisotropy parameter, and $d$ is the interlayer distance. 
The distance $a$ between Josephson vortices in the same 
layer and the period $N$ (see Fig.\ \ref{fig1}(a)) are 
related with the in-plane field along the $x$ axis $B_{x}$ 
as $\phi_{0}=adNB_{x}$. Since the ground state should 
be uniform along the field direction, the free energy 
functional per unit volume is expressed as
\begin{eqnarray}
f\{\phi_{n}(\bar{y})\}
&=&\frac{\phi_{0}B_{x}}{16\pi^{2}\gamma\lambda_{ab}^{2}}u\{\phi_{n}(\bar{y})\},\\*
u\{\phi_{n}(\bar{y})\}
&=&\frac{1}{\pi}\sum_{n=1}^{N}\int_{0}^{\bar{a}}d\bar{y} \left[\frac{1}{2}
                \left(\frac{d\phi_{n}}{d\bar{y}}\right)^{2}\right.\nonumber\\*
&+&\Biggl.1-\cos\left(\phi_{n+1}-\phi_{n}-h\bar{y}\right)\Biggr],
\label{scale-u}
\end{eqnarray}
where the following normalized quantities are introduced:
\begin{equation}
\label{field}
    \bar{y}\equiv\frac{y}{\gamma d}\ ,\ \ 
    \bar{a}\equiv\frac{a}{\gamma d}\ ,\ \ 
          h\equiv\frac{2\pi}{\phi_{0}}\gamma d^{2}B_{x}.
\end{equation}
By minimizing the free energy functional (\ref{scale-u}) 
with respect to $\phi_{n}$, we have
\begin{equation}
\frac{d^{2}\phi_{n}}{d\bar{y}^{2}}
+\sin(\phi_{n+1}-\phi_{n}-h\bar{y})
-\sin(\phi_{n}-\phi_{n-1}-h\bar{y})=0.
\end{equation}
This differential equation is solved in the unit cell $\bar{a}\times Nd$ 
(see Fig.\ \ref{fig1}(a)) with appropriate quasi periodic boundary 
conditions \cite{Koshelev05}. In order to search the ground state, 
the shift from the aligned vortex lattice $\Delta$ defined in 
Fig.\ \ref{fig1}(a) should also be swept as the in-plane field $h$.

Another description of the vortex lattice in layered material is 
shown in Fig.\ \ref{fig1}(b), where the vector ${\bf \bar{a}}$ 
between the nearest-neighbor vortices in the same layer and the 
unit vectors of the triangular vortex lattice ${\bf e_{1}}$ and 
${\bf e_{2}}$ are combined with a pair of integers $n$ and $m$ 
as ${\bf \bar{a}}=n{\bf e_{1}}+m{\bf e_{2}}$. This labeling is 
based on an equilateral triangular vortex lattice, which we call 
the ``rigid rotated vortex lattice" in the present article. 
For each $(n,m)$, there is a corresponding fractional value of 
$\Delta$. However, it is not trivial that any rigid rotated 
vortex lattice can be a ground state. As will be revealed 
below, it is not the case. Hereafter the energy function 
$g\equiv u+\frac{1}{2}\log h-1.432$ is used instead of $u$, 
following Ref.\ \cite{Koshelev06}.

\begin{figure}
\begin{center}
\caption{\label{fig2}(a) Energy function $-g$ versus the shift from the 
aligned vortex lattice $\Delta$ and the normalized in-plane field $h$. 
Each color stands for the energy function for each $N$ [color online]. 
The dark [blue] solid lines denote the ground state, and the pale [green] 
solid lines stand for the lowest-energy state (but not the ground state) 
for each $N$. The rotated-lattice index $(n,m,N)$ for each ground state 
configuration is displayed. (b) Top view of the above figure 
[gif figure appended].
}
\end{center}
\end{figure}
\begin{figure}
\begin{center}
\hspace*{-4mm}
\includegraphics[width=87mm]{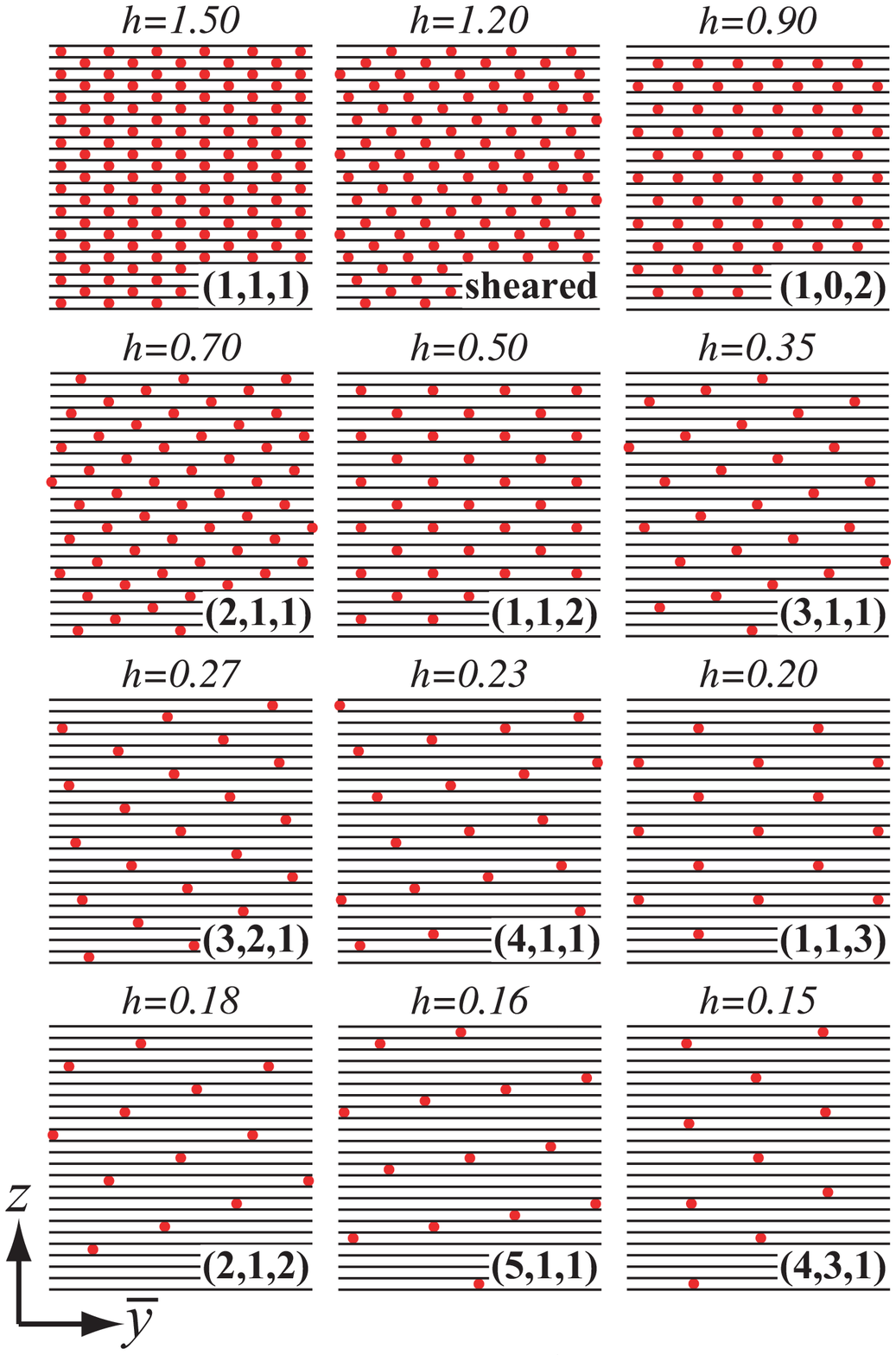}
\caption{\label{fig3}Ground-state configurations for various in-plane 
fields $h$ with the corresponding rotated-lattice indices ($n,m,N$).}
\end{center}
\end{figure}
\medskip
\par
{\it Results.}
The energy landscape of eq.\ (\ref{scale-u}) is shown in Fig.\ \ref{fig2}(a), 
where $-g$ is taken instead of $g$ because the valley structure is hard to 
visualize. The ridges in this figure correspond to the valleys in the actual 
energy landscape. Energy functions for various $N$'s are drawn in the same 
figure, and different colors correspond to different $N$'s. The ground 
state denoted by dark solid lines often jumps between two different values 
of $\Delta$ as drawn in dashed lines. The pale solid lines represent the 
lowest-energy state (but not the ground state) for each $N$.

For high fields, the aligned $(1,1,1)$ vortex lattice corresponds to the 
ground state with $\Delta=0$. As the field decreases to $h\approx 1.33$, 
a sheared lattice with finite $\Delta$ becomes the ground state. The 
shearing angle increases continuously as the field decreases, and such 
continuous field dependence cannot be represented by the rotated-lattice 
index. Although the ground state changes to the $(1,0,2)$ aligned lattice 
at $h\approx 0.99$, the sheared vortex lattice state is still the lowest 
energy state in the $N=1$ subspace, and it becomes the ground state 
again at $h\approx 0.80$. The field dependence of $\Delta$ is quite 
small around $h\approx 0.7$, which corresponds to the simplest rigid 
rotated vortex lattice labeled with $(2,1,1)$ and characterized by 
$\Delta_{\rm rigid}=1/7$. Since each ground-state configuration is 
separated in the energy landscape, these configurations can be 
labeled by the rotated-lattice indices $(n,m,N)$, even though 
the parameter $\Delta$ may deviate from the fractional values 
$\Delta_{\rm rigid}$ corresponding to the rigid rotated vortex lattices. 
Typical ground-state configurations are shown in Fig.\ \ref{fig3}, where 
all the rotated-lattice indices $(n,m,N)$ for $h\geq 0.15$ are listed.

\begin{figure}[t]
\begin{center}
\caption{\label{fig4}Energy function $-g$ versus $\Delta$ and $h$ in 
the $N=1$ subspace [color online]. The dark [blue] or pale [green] 
solid lines are the same as in Fig.\ \ref{fig2}(a), and the paler 
[yellow] solid lines stand for the local energy minima (not the 
lowest-energy states), and its top view is displayed in the inset. 
The rotated-lattice indices are given for all the ground-state 
configurations in this subspace [gif figure appended].}
\end{center}
\end{figure}
Next, the energy landscape in the $N=1$ subspace is shown in 
Fig.\ \ref{fig4}. Here the dark or pale solid lines are defined 
similarly to the ones in Fig.\ \ref{fig2}(a), and the paler solid 
lines denote the local energy minima (not the lowest-energy state 
in this subspace). Apparently, ground-state configurations of the 
rotated lattices are connected with each other by lines of local 
energy minima, and such ground-state configurations correspond 
to the bifurcation points of local minima. Similar bifurcating 
structure is expected for $N\geq 2$ at much low fields.

\begin{figure}[t]
\begin{center}
\includegraphics[width=88mm]{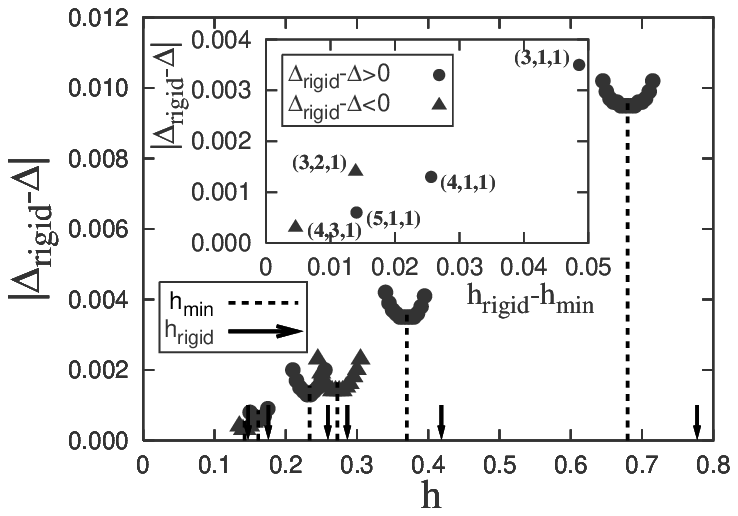}
\caption{\label{fig5}Deviation of $\Delta$ from that of the rigid 
rotated vortex lattice ($\Delta_{\rm rigid}$) versus normalized 
field $h$ in the $N=1$ subspace, and it takes minimum at 
$h=h_{\rm min}$ (dashed lines). The fields corresponding 
to the rigid rotated vortex lattices ($h_{\rm rigid}$) are 
marked by arrows. In the inset, the deviation of $\Delta$ 
($|\Delta_{\rm rigid}-\Delta|$) at $h=h_{\rm min}$ is 
plotted versus that of $h$ ($h_{\rm rigid}-h_{\rm min}$) 
together with the rotated-lattice indices.
}
\end{center}
\end{figure}
From precise analyses of $\Delta$, we find that the rigid rotated vortex 
lattices expected from the London theory can never be the ground state. 
The deviation of $\Delta$ from the values corresponding to the 
rigid rotated lattices $\Delta_{\rm rigid}$ is plotted versus $h$ 
in Fig.\ \ref{fig5}. The inequality $\Delta_{\rm rigid}-\Delta>0$ 
holds for $n\geq m/2$, and vice versa. The field corresponding to 
the rigid rotated vortex lattice $h_{\rm rigid}$ is marked by 
arrows in the same figure, and $|\Delta_{\rm rigid}-\Delta|$ 
takes minima at $h=h_{\rm min}$ (dashed lines). The deviation 
decreases as $h$ decreases, and seems to vanish in the $h\to 0$ 
limit. In the inset, the deviation of $\Delta$ at $h=h_{\rm min}$ 
is plotted versus that of the field, $h_{\rm rigid}-h_{\rm min}$.

\begin{figure}[t]
\begin{center}
\includegraphics[width=88mm]{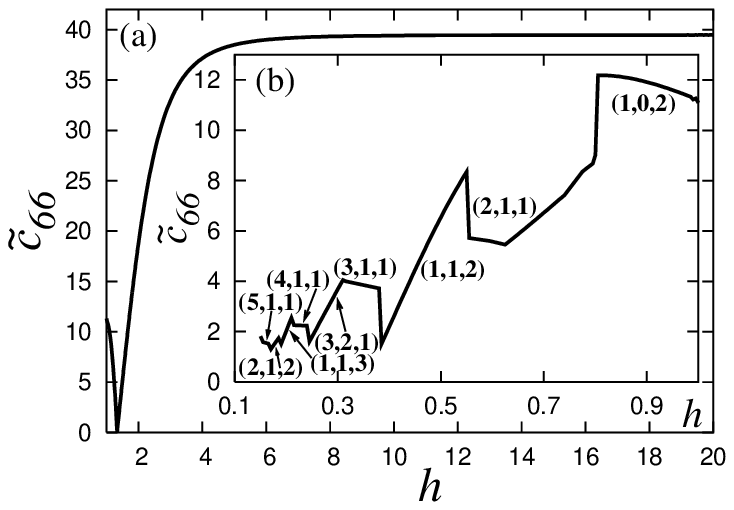}
\caption{\label{fig6}Field dependence of the scaled shear modulus 
$\tilde{c}_{66}$ defined in eq.\ (\ref{c66}). (a) For $h\geq 1$, 
$\tilde{c}_{66}$ converges to a constant value $C\approx 39.4$ for 
high fields, and the vanishing point of $\tilde{c}_{66}$ divides 
the aligned and sheared vortex lattices. (b) For $h \leq 1$, 
corresponding rotated-lattice indices are displayed.
}
\end{center}
\end{figure}
Finally, we observe the shear modulus $c_{66}$ defined by
\begin{equation}
  \label{c66}
  c_{66}\equiv \frac{d^{2}N^{2}}{a^{2}}\frac{\partial^{2}f}{\partial \Delta^{2}}
        \equiv\frac{\phi_{0}^{2}}{(8\pi\lambda_{ab})^{2}}
              \frac{\tilde{c}_{66}}{2\pi^{3}\gamma^{4}d^{2}},\ 
  \tilde{c}_{66}=N^{4}h^{3}\frac{\partial^{2}u}{\partial\Delta^{2}},
\end{equation}
where $\phi_{0}=adNB_{x}$ and eq.\ (\ref{field}) are substituted and 
parameter dependence on $h$, $N$ and $\Delta$ is only included in the scaled 
shear modulus $\tilde{c}_{66}$. Field dependence of $\tilde{c}_{66}$ 
is displayed in Fig.\ \ref{fig6}, and this quantity converges to a 
constant value $C\approx 39.4$ in the $h\to\infty$ limit. Namely, 
the shear modulus becomes field independent for high fields, 
\begin{equation}
  \label{c66sc}
  c_{66}\approx \frac{\phi_{0}^{2}}{(8\pi\lambda_{ab})^{2}}
                \frac{C}{2\pi^{3}d^{2}}\gamma^{-4}.
\end{equation}
This behavior is in sharp contrast to the exponentially vanishing 
$c_{66}$ of the London model \cite{Ivlev}, and is also different 
from that of the elastic theory, 
$c_{66}=[\phi_{0}B_{x}/\left(8\pi\lambda_{ab}\right)^{2}]\gamma^{-3}$ \cite{Kogan89}. 
Introducing a ``saturated field" $\tilde{B}$ defined below, eq.\ (\ref{c66sc}) 
can be expressed similarly to the elastic theory as
\begin{equation}
  c_{66}\approx\frac{\phi_{0}\tilde{B}^{x}}{\left(8\pi\lambda_{ab}\right)^{2}}\gamma^{-3},\ 
  \tilde{B}^{x}=\frac{C\phi_{0}}{2\pi^{3}\gamma d^{2}}, 
\end{equation}
and $\tilde{a}\equiv\phi_{0}/(d\tilde{B}^{x})=(2\pi^{3}/C)\gamma d \approx 1.57\gamma d$ 
corresponds to the effective core size of a Josephson vortex.
As the field decreases, $\tilde{c}_{66}$ vanishes at the shear 
instability field of the aligned vortex lattice \cite{Ivlev}, 
increases again together with the increase of the shearing 
angle, and jumps when the rotated-lattice index changes.

\medskip
\par
{\it Discussion.}
In the London model, only the rigid vortex lattices can be the 
ground state by definition. Therefore, the sheared vortex lattices 
are specific to the models including the layered structure explicitly 
such as the present model. We find that the rotated vortex lattices 
with $\Delta_{\rm rigid}$ cannot be the ground state even at the 
fields corresponding to the rigid rotated lattices, which is 
another nontrivial effects of the layered structure.

It would be interesting to investigate how the above description is 
altered by various kinds of fluctuations. Hu {\it et al.} \cite{Hu05} 
studied the phase diagram of interlayer Josephson vortex systems in the 
vicinity of the melting temperature using the density functional theory. 
They found the vortex smectic state around the $(n,m,N)=(1,0,2)$ 
state, while no sheared vortex lattice states were observed between 
the $(1,1,1)$ and $(1,0,2)$ states. This finding may suggest that 
the sheared vortex lattice state is affected very much by thermal 
fluctuations. Study on temperature dependence of the deviation from 
the rigid rotated lattices would be an interesting future problem.

The screening effect is also studied in the present framework, and 
we find that the rotated vortex lattices are almost unchanged, while 
the sheared vortex lattices or the $N\ge 2$ aligned vortex lattices 
are affected very much. These facts suggest that the rotated vortex 
lattice structure is robust against the screening effect. Details 
of this study will be reported elsewhere \cite{Nonomura06}.

\medskip
\par
{\it Summary.}
Ground state of interlayer Josephson vortex systems is investigated 
on the basis of the simplified Lawrence-Doniach model, in which 
spatial dependence of the gauge field and the amplitude of 
superconducting order parameter is neglected. The free energy 
functional is minimized in the unit cell with one Josephson 
vortex. We draw the energy landscape with respect to the 
in-plane field, the period of insulating layers including 
Josephson vortices, and the shift from the aligned lattice.

As the in-plane field decreases, the aligned vortex lattice 
along superconducting layers changes to the sheared vortex 
lattice and then to the rotated vortex lattices, in which 
the shift from the aligned lattice is approximately given 
by that of the corresponding rigid rotated lattices. 
Systematic deviation from the rigid configurations is 
clarified for the first time. Ground-state configurations 
of the rotated lattices are connected with each other by 
local minima in the energy landscape to form a multi-valley 
structure, and such ground-state configurations correspond 
to the bifurcation points of the valleys.

In the high-field region, the shear modulus becomes 
independent of field and its anisotropy dependence is given by 
$c_{66}\propto \gamma^{-4}$, quite different from the London 
theory in which $c_{66}$ decays exponentially with field. 
This quantity vanishes at the onset of the sheared vortex 
lattice, and jumps when the rotated-lattice index changes.

\medskip
\par
{\it Acknowledgement.}
We would like to thank Dr.\ A.~E.~Koshelev for helpful discussions 
and comments, and sending his articles prior to public appearance.


\begin{thebibliography}{99}
\bibitem{Ivlev}
B.~I.~Ivlev {\it et al.}, J.\ Low Temp.\ Phys.\ {\bf 80}, 187 (1990).
%B.~I.~Ivlev, N.~B.~Kopnin, and V.~L.~Pokrovsky, 
%J.\ Low Temp.\ Phys.\ {\bf 80}, 187 (1990).
\bibitem{Koshelev06}
A.~E.~Koshelev, cond-mat/0602341.
\bibitem{Campbell}
L.~J.~Campbell {\it et al.}, Phys.\ Rev.\ B {\bf 38}, 2439 (1988).
%L.~J.~Campbell, M.~M.~Doria, and V.~G.~Kogan, 
%Phys.\ Rev.\ B {\bf 38}, 2439 (1988).
\bibitem{Levitov}
L.~S.~Levitov, Phys.\ Rev.\ Lett.\ {\bf 66}, 224 (1991).
\bibitem{Lagna}
M.~F.~Laguna {\it et al.}, Phys.\ Rev.\ B {\bf 62}, 6692 (2000).
%M.~F.~Laguna, D.~Dom\'inguez, C.~A.~Balseiro, 
%Phys.\ Rev.\ B {\bf 62}, 6692 (2000).
\bibitem{Bulaevskii91}
L.~Bulaevskii and J.~R.~Clem, Phys.\ Rev.\ B {\bf 44}, 10234 (1991).
\bibitem{Ichioka95}
M.~Ichioka, Phys.\ Rev.\ B {\bf 51}, 9423 (1995).
\bibitem{Hu98}
X.~Hu and M.~Tachiki, Phys.\ Rev.\ Lett.\ {\bf 80}, 4044 (1998).
\bibitem{Ikeda02}
R.~Ikeda, J.\ Phys.\ Soc.\ Jpn.\ {\bf 71}, 587 (2002).
\bibitem{Koshelev05}
A.~E.~Koshelev, Private Communication.
\bibitem{Kogan89}
V.~G.~Kogan and L.~J.~Campbell, Phys.\ Rev.\ Lett.\ {\bf 62}, 1552 (1989).
\bibitem{Hu05}
X.~Hu {\it et al.}, Phys.\ Rev.\ B {\bf 72}, 174503 (2005).
%X.~Hu, M.-B.~Luo, and Y.-Q.~Ma, Phys.\ Rev.\ B {\bf 72}, 174503 (2005).
\bibitem{Nonomura06}
Y.~Nonomura and X.~Hu, in preparation.
\end{thebibliography}
\end{document}